\documentclass[11pt, A4paper]{article} 

\usepackage[ansinew]{inputenc}
\usepackage{amsmath, amssymb, graphics, amsthm}
\usepackage{epsfig}
\usepackage{color, xcolor}
\usepackage{fancyhdr} 
\usepackage{manfnt}
\usepackage[T1]{fontenc}
\usepackage{hyperref}

\usepackage[normalem]{ulem}

\newcommand{\mN}{\mathcal N}

\newcommand{\ket}[1]{\left| #1 \right\rangle}

\makeatletter
\@addtoreset{equation}{section}
\makeatother

\newcommand{\be}{\begin{equation}}
\newcommand{\ee}{\end{equation}}
\newcommand{\ba}{\begin{eqnarray}}
\newcommand{\ea}{\end{eqnarray}}

\def\pb#1{\rlap{\lower1.5ex\hbox{$\longleftarrow$}}{#1}}
\def\dpb#1{\rlap{\lower1.5ex\hbox{$\Longleftarrow$}}{#1}}
\def\spb#1{\rlap{\lower1.0ex\hbox{$\leftarrow$}}{#1}}
\def\sdpb#1{\rlap{\lower1.0ex\hbox{$\Leftarrow$}}{#1}}

\title{{\sf A note on quantum supergravity and AdS/CFT}}
\author{
{\sf N. Bodendorfer}\thanks{{\sf 
norbert.bodendorfer@fuw.edu.pl}}\\
\\
{\sf  Faculty of Physics, University of Warsaw, Pasteura 5, 02-093, Warsaw, Poland}\\
}
\date{{\small\sf \today}}

\begin{document} 

\maketitle

{\sf

\begin{abstract}

We note that the non-perturbative quantisation of supergravity as recently investigated using loop quantum gravity techniques provides an opportunity to probe an interesting sector of the AdS/CFT correspondence, which is usually not considered in conventional treatments. In particular, assuming a certain amount of convergence between the quantum supergravity sector of string theory and quantum supergravity constructed via loop quantum gravity techniques, 
we argue that the large quantum number expansion in loop quantum supergravity corresponds to the $1/N^2_c$ expansion in the corresponding gauge theory. In order to argue that we are indeed dealing with an appropriate quantum supergravity sector of string theory, high energy ($\alpha'$) corrections are being neglected, leading to a gauge theory at strong coupling, yet finite $N_c$. 
The arguments given in this paper are mainly of qualitative nature, with the aim of serving as a starting point for a more in depth interaction between the string theory and loop quantum gravity communities. 

\end{abstract}

}

\section{Introduction}

Finding a theory of quantum gravity is considered as one of the main open problems in contemporary theoretical physics. Diverse approaches to the subject have been proposed \cite{OritiApproachesToQuantum}, however little convergence between them has been observed so far. The main problem of quantum gravity research is certainly missing experimental input, which forces researchers to rely on criteria such as internal consistency or, more subjectively, mathematical elegance. While cosmological observations related to physics of the very early universe arguably offer the most probable scenario for detecting quantum gravity effects in the near future, they will probably not fully settle the issue of which approach to quantum gravity is the correct one. 

In this situation, it is very desirable to find alternative applications for the tools and theories developed in the context of quantum gravity. Within string theory \cite{GreenBook1, GreenBook2, PolchinskiBook1, PolchinskiBook2}, the AdS/CFT correspondence \cite{MaldacenaTheLargeN, WittenAntiDeSitter, GubserGaugeTheoryCorrelators} has been highly successful exactly along these lines: a conjectured, but well motivated equivalence between a gravitational (string) theory on the one hand, and a gauge theory on the other hand, has been used to gain important insights into gauge theories at strong coupling, which for example govern QCD or condensed matter physics, see e.g. \cite{AmmonGaugeGravityDuality} and references therein. 

When performing concrete computations in AdS/CFT, one usually takes certain limits, in which the string theory side is consecutively approximated by tree-level (``classical'') string theory and then by classical supergravity. In turn, the dual gauge theory is considered in the limit of a large number of colours and at strong 't Hooft coupling. Corrections to this limit can for example be included by computing the respective correction terms to the low energy effective action, which is just supergravity if all corrections are neglected. While many important insights were gained in this limit, it is still very desirable to probe also different regimes of the AdS/CFT correspondence. In particular, a sector corresponding to gauge theories with a finite number of colours and strong coupling is very desirable, since this is (up to supersymmetry) also the case in nature. Such a sector is however currently unavailable using standard AdS/CFT techniques, mainly because string theory is non-perturbative in this context and thus very challenging. 

The main purpose of this paper is to point out that using techniques from loop quantum gravity \cite{RovelliQuantumGravity, ThiemannModernCanonicalQuantum, RovelliBook2}, a quantisation of supergravity has been constructed \cite{BTTVIII} which is a good candidate to describe string theory in the appropriate limit corresponding to a strongly coupled gauge theory with a finite number of colours. While there is so far no quantitative evidence for this assertion, we provide a heuristic argument in this paper that this might indeed be the case. It is our hope that the potential benefits, to access a new and particularly interesting sector of string theory and apply it to the AdS/CFT correspondence, will motive more in depth research in this topic. In particular, we hope to spark some exchange, interaction and collaboration between researchers from string theory, AdS/CFT, and loop quantum gravity. Having this in mind, this paper targets a general audience and technicalities are kept to a minimum.

\section{Quantum supergravity and AdS/CFT}

Maldacena's original proposal for a gauge / gravity correspondence relates type IIB string theory on AdS${}^5 \times S^5$ and $\mN = 4$ Super Yang-Mills theory. 
Both sides of the correspondence are governed by two free parameters, the string coupling $g_s$ and the string length $l_s = \sqrt{\alpha'}$ on the gravity side, and the number of colours $N_c$ and the 't Hooft coupling $\lambda = g_{\text{YM}}^2 N_c$ on the gauge theory side. 
The correspondence can be stated in different forms, depending on which limits of the involved theories are considered. The strongest possible statement is an exact equivalence between the two theories, that is for arbitrary string coupling $g_s$ and string length $l_s$. The correspondence then says that the partition functions of the two theories agree, 
\be
	Z_{\mN=4} = Z_{\text{AdS}^5 \times S^5}
\ee
and that the free parameters are related as
\be
	N^2_c \propto \frac{L^8}{G_{10}}, ~~~~ \lambda \propto \left(\frac{L}{l_s}\right)^4 \text{,} \label{eq:NLambda}
\ee
where $L$ is the radius of the anti-de Sitter space and $G_{10} \propto g_s^2 l_s^8$ is the induced $10$-dimensional Newton constant in the supergravity low energy effective action of string theory. 

It is in general very hard to verify the AdS/CFT correspondence in the above, most general form, since computations in both string and gauge theory are hard for generic values of the coupling constants. Meanwhile, a large amount of evidence for the validity of the conjecture has been accumulated in certain limiting cases of the correspondence. A first simplification is obtained in the limit $N_c \rightarrow \infty$, while keeping $\lambda$ constant. On the gauge theory side, this corresponds to the 't Hooft limit where only planar diagrams contribute. On the string theory side, only tree diagrams contribute in string perturbation theory. A second simplification on the string theory side arises if one furthermore considers the limit $l_s / L \rightarrow 0$, corresponding to $\lambda \rightarrow \infty$, i.e. the classical supergravity approximation of string theory. In practise, most computations on the gravity side are performed in this limit, in turn yielding insights into the strongly coupled large $N_c$ regime of the corresponding gauge theory. 
 
 While computations performed in the above regime have lead to many interesting insights into strongly coupled systems, see e.g. \cite{AmmonGaugeGravityDuality} and references therein, they do not have to agree with observation since they are valid only in the large $N_c$ limit\footnote{We neglect here the issue of supersymmetry, which so far has not been observed in nature.}. Keeping $N_c$ fixed on the other hand, we are free to vary the 't Hooft coupling $\lambda$ and send it to infinity, corresponding to a negligible string length $l_s$ as compared to the AdS radius $L$. In this limit, massive string states which are not included in the supergravity description can be neglected and quantum supergravity should be a valid approximation on the string theory side. Since
\be
	g_s  \propto \frac{\lambda}{N_c} \text{,}
\ee
this limit corresponds to the strong coupling limit of string theory at vanishing string scale (keeping $N_c = L^4/ g_s l_s^4$ fixed), which is in particular not accessible by standard perturbative means. 
It is thus strongly desirable to have a theory of quantum supergravity at one's disposal which is well defined for generic values of $L^8/G_{10}$ and preferably also allows an expansion in this parameter (as opposed to $g_s$). Also, quantum corrections should be governed by the magnitude of $L^8/G_{10}$, whereas the limit $L^8/G_{10} \rightarrow \infty$ should correspond to classical supergravity. 
{\it The central message of this note is to point out that the recently proposed quantisation of diverse supergravity theories using methods of loop quantum gravity \cite{BTTVIII, BTTI, BTTII, BTTIII, BTTIV, BTTXII, BTTV, BTTVI, BTTVII} provides such a proposal:}

At the classical level, $d$-dimensional Lorentzian general relativity in Hamiltonian form can be reformulated as an SO$(d)$ gauge theory, i.e. as a Yang-Mills theory with compact gauge group and subject to several constraints. Using this starting point, quantisation techniques of loop quantum gravity can be applied, resulting in a kinematical Hilbert space given by the square integrable functions on the space of (generalised) connections with respect to the so-called Ashtekar-Lewandowski measure \cite{AshtekarRepresentationsOfThe, AshtekarRepresentationTheoryOf}. Dynamics can be defined by regularising constraint operators following techniques proposed by Thiemann \cite{ThiemannQSD1}. Matter fields, in particular those of many interesting supergravity theories, were also incorporated into the formalism. Through the use of the internal gauge group, representation theory of SO$(d)$ enters the quantum theory. In particular, the so called simplicity constraint, which enters the theory already at the classical level in order to ensure equivalence to general relativity (or supergravity), enforces that only a certain sub-class of representations are allowed. These representations are labelled by a single non-negative integer\footnote{In the original literature such as \cite{BTTVIII}, the representation label is denoted by $\lambda$, which however conflicts with the 't Hooft coupling in this article.} $l$. The quantum number $l$ most prominently enters the spectrum of the spacelike $(d-2)$ area operator $\hat A$, whose action on a holonomy with representation label $l$ (i.e. a Fock-like geometric excitation in the Hilbert space) is given by 
\be
	\hat A \ket{l} \propto G_{d} \sqrt{l (l + d-2)} \ket{l} \text{.}
\ee
Specialising to $d=10$, we see that relative geometric scales such as $L^8 / G_{10}$, with $G_{10} \sim l_p^8$, are determined by the scale of the involved quantum numbers $l$. 
In fact, the cosmological scale $L$ sets the scale of the geometry in AdS${}^5$ (and similarly in the $S^5$ factor), which in static slicing and with the redefinition $\hat r = r/L$ reads\footnote{We present here only the metric in the Poincar\'e patch, where AdS/CFT is best understood and the boundary is flat Minkowski space. The same argument however also holds in global coordinates.}

\be
			ds^2 = - \hat r^2 dt^2 + L^2 \left( \frac{d \hat r^2}{\hat r^2}+ \hat r^2 d \Omega^2 \right) \text{.}
\ee
We note that the infinite volume of the spatial slice can be incorporated into a loop quantum gravity type quantisation using Penrose's idea of conformal compactification \cite{PenroseAsymptoticPropertiesOf}, see \cite{BConformalCompactificationLQG}. The scaling of the compactified {\it finite} spatial geometry with $L$ still holds in this case. 

In other words, in the process of encoding a slice of anti-de Sitter space in a cylindrical function over a graph (a wave function in loop quantum gravity), the anti-de Sitter radius $L$ is determined by the number of edges and vertices in the graphs as well as the magnitude of the involved spins\footnote{There is another free parameter of the theory entering the spectrum of geometric operators, known as the Barbero-Immirzi parameter in case of $SU(2)$ connection variables. Whether or not this parameter exhibits some sort of running in the context of a classical limit of the theory is still matter of debate, see e.g. \cite{BenedettiPerturbativeQuantumGravity, BNI, HanCovariantLoopQuantum}.}. Keeping the graph fixed and scaling the quantum numbers thus corresponds to scaling the ratio $L^8 / G_{10}$. Such a scaling is of course well known within loop quantum gravity, see e.g. \cite{ConradySemiclassicalLimitOf, BarrettLorentzianSpinFoam, HanOnSpinfoamModels}, and mostly referred to as the large spin approximation in the context of $3$ and $4$ dimensions, where the gauge group can be taken to be SU$(2)$ \footnote{The related cutoff for the quantum numbers induced by the cosmological constant is also familiar from spin foam models, where a positive cosmological constant can be incorporated by using quantum groups \cite{Han4DimensionalSpin, FairbairnQuantumDeformationOf}. A finite cosmological constant, through the quantum number cutoff, thus also provides a cutoff on how ``classical'' the theory can behave.}. In the large quantum number limit, which is considered in loop quantum gravity (in a certain precise sense) as the classical limit, we would thus presumably\footnote{While a large amount of evidence has been accumulated that general relativity is indeed the classical limit of $4$-dimensional loop quantum gravity, see e.g. \cite{GieselAQG2, HanCovariantLoopQuantum} and references therein, no such investigation has been performed in higher dimensions so far.} recover (a lattice truncated version of) classical supergravity, in accordance with first taking the $g_s \rightarrow 0$ limit at fixed $\lambda$, and then sending $\lambda \rightarrow \infty$ as in the standard treatment, also resulting in a classical supergravity approximation on the string theory side of the correspondence.

\section{Comments}

It is important to stress that on the loop quantum gravity side, there are still several open issues to be resolved before generic calculations can be performed in the context of AdS/CFT. For example, an understanding of anti-de Sitter space and its asymptotic symmetries is mandatory. First steps along this route were taken in \cite{BConformalCompactificationLQG}, as well as in the asymptotically flat context in \cite{ThiemannQSD6, VaradarajanAQuantumKinematics}. In particular, the issue of the emergence of a spacetime continuum via coarse graining is directly related to the question of determining asymptotic symmetries. In practise however, one would expect to first obtain a discretised version of the boundary field theory without taking the continuum limit in the bulk and try to establish a correspondence at this level.

While the $3$-form fields of $11$-dimensional supergravity have been successfully incorporated into a loop quantisation \cite{BTTVII}, the self-duality condition on the $4$-form gauge field in type IIB supergravity has not been addressed yet. This issue can however be circumvented by compactifying the $S^5$ and working with the resulting $5$-dimensional supergravity\footnote{The methods for the quantisation of Majorana fermions developed in \cite{BTTVI} however do not apply to $5$ dimensions. Still, in many practical calculations, fermions are ignored and we can work with the bosonic sector of the theory instead.}. 

Also, the issue of local supersymmetry has not been fully addressed so far within loop quantum supergravity. While the supersymmetry constraint, the generator of supersymmetry transformations, has been quantised using similar methods as for the Hamiltonian constraint \cite{BTTVI}, its quantum algebra and action on quantum states have not been studied so far. In particular, the issue of possible anomalies needs to be understood. 

In order to go beyond the quantum supergravity approximation of string theory, $\alpha'$-corrections, i.e. higher curvature corrections, can be added to the supergravity action. Taking these into account within a loop quantisation then offers a chance improve on the quantum supergravity approximation of full string theory.

Another limit of string theory worth considering in the context of a loop quantum gravity type quantisation and AdS/CFT is $l_s /L \rightarrow \infty$, corresponding to higher spin gravity \cite{VasilievHigherSpinGauge}. In particular, the formulation of (truncated) $3$-dimensional higher spin gravity as an SL$(N, \mathbb R) \times$ SL$(N, \mathbb R)$ Chern-Simons theory or an SL$(N, \mathbb R)$ BF-theory \cite{CampoleoniAsymptoticSymmetriesOf} suggests to apply methods from $3$-dimensional loop quantum gravity. 

In the context of three dimensions, it is also important to mention \cite{FreidelReconstructingAdSCFT}, which discusses quantum gravity aspects of the AdS/CFT correspondence in three dimensions. Since quantum gravity is currently best understood there, this line of research is very important and needs to be developed further, see e.g. \cite{CianfraniWheelerDeWittEquation} and the reports \cite{LivineTBP, Dittrich3DHolographyFromTalk} about ongoing work. It is also interesting to note that in the context of AdS$_3$/CFT$_2$, the large quantum number limit in loop quantum gravity, here the large spin limit due to the gauge group SU(2) \cite{ThiemannQSD4}, corresponds to the large central charge limit due to $j \sim L / G_3 \sim c$, which is again a semiclassical limit in the CFT.

\section{Conclusion}

In this paper, we have pointed out that there is a strong potential for the methods developed within loop quantum gravity to be useful for the AdS/CFT correspondence, and thus also for string theory. In particular, the non-perturbative quantisation techniques of loop quantum gravity allow to formulate a theory of quantum supergravity that could be used to make computations within AdS/CFT. The regime in which this proposal was argued to be valid is that of a finite number of colours $N_c$ in the dual gauge theory, whereas the 't Hooft coupling $\lambda$ is taken to be large in order to be able to neglect stringy effects. 

While it was pointed out that there are several technical and conceptual issues to be dealt with on the loop quantum gravity side before generic computations can be performed, the large potential benefits of this line of research clearly outweigh the risks: 1) Probing strongly coupled gauge theories at a finite number of colours is of significant interest to both elementary particle and condensed matter physics. 2) Gaining a deeper understanding of the AdS/CFT correspondence beyond classical supergravity limit is of significant interest to theoretical and mathematical physics. 3) Understanding the relation between string theory and loop quantum gravity as the largest approaches to quantum gravity is strongly desirable to quantum gravity researchers given the scarcity of experimental data. In particular, if the AdS/CFT predictions made via loop quantum supergravity fail to be correct, it will nevertheless be interesting to understand whether this means that there is a problem with AdS/CFT, or that the quantum supergravity sector of string theory is not described by a non-perturbative quantisation of supergravity. 
With this in mind, we hope that the prospects of making progress with the above issues will draw some attention to the topic discussed in this paper.

\section*{Acknowledgements}
NB was supported by a Feodor Lynen Research Fellowship of the Alexander von Humboldt-Foundation and during final improvements of this manuscript by the grant of Polish Narodowe Centrum Nauki nr 2012/05/E/ST2/03308. 
Many discussions with Martin Ammon, Andreas Sch\"afer, and John Schliemann are gratefully acknowledged. This paper summarises a talk given in January 2015 at the University of Regensburg, during the preparation of which the main argument was conceived.

\bibliographystyle{utphysmendeley}
\bibliography{library}

\end{document}